\documentclass[prl,reprint,superscriptaddress,showpacs,preprintnumbers,amsmath,amssymb, showkeys, hidelinks, nofootinbib]{revtex4-1}

\pdfoutput=1

\usepackage{amsmath}
\usepackage{graphicx}
\usepackage{natbib}
\usepackage{verbatim}
\usepackage[withbib, all]{authorindex}	
\usepackage[usenames]{color}
\usepackage{bm}
\usepackage{hyperref}

\hypersetup{
    colorlinks, linkcolor={blue},
    citecolor={blue}, urlcolor={blue},
}

\begin{document}

\title{Off-Resonant Manipulation of Spins in Diamond via Precessing Magnetization of a Proximal Ferromagnet}

\author{C. S. Wolfe*}
\affiliation{Department of Physics, The Ohio State University, Columbus, Ohio 43210, USA}

\author{V. P. Bhallamudi*}
\affiliation{Department of Physics, The Ohio State University, Columbus, Ohio 43210, USA}
\let\thefootnote\relax\footnote{* C. Wolfe and V. Bhallamudi contributed equally to this work.}

\author{H. L. Wang}
\affiliation{Department of Physics, The Ohio State University, Columbus, Ohio 43210, USA}

\author{C. H. Du}
\affiliation{Department of Physics, The Ohio State University, Columbus, Ohio 43210, USA}

\author{S. Manuilov}
\affiliation{Department of Physics, The Ohio State University, Columbus, Ohio 43210, USA}

\author{ A. J. Berger}
\affiliation{Department of Physics, The Ohio State University, Columbus, Ohio 43210, USA}

\author{R. Adur}
\affiliation{Department of Physics, The Ohio State University, Columbus, Ohio 43210, USA}

\author{F. Y. Yang}
\affiliation{Department of Physics, The Ohio State University, Columbus, Ohio 43210, USA}

\author{P. C. Hammel}\email{hammel@physics.osu.edu}
\affiliation{Department of Physics, The Ohio State University, Columbus, Ohio 43210, USA}

\date{\today}

\begin{abstract}
We report the manipulation of nitrogen vacancy (NV) spins in diamond when nearby ferrimagnetic insulator, yttrium iron garnet, is driven into precession. The change in NV spin polarization, as measured by changes in photoluminescence, is comparable in magnitude to that from conventional optically detected magnetic resonance, but relies on a distinct mechanism as it occurs at a microwave frequency far removed from the magnetic resonance frequency of the NV spin. This observation presents a new approach to transferring ferromagnetic spin information into a paramagnet and then transducing the response into a robust optical signal. It also opens new avenues for studying ferromagnetism and spin transport at the nanoscale.
\end{abstract}

\pacs{72.25.Mk, 75.76.+j, 75.78.-n, 75.78.-n, 75.78.-n}
\keywords{NV, magnetization dynamics, FMR, spin transport, YIG}
\maketitle

Understanding the transport of spin and energy between dissimilar materials is a topic of intense current interest reflecting both the scientific richness of the topic as well as it technological potential\cite{slichter, meier_optical_1984,Hammel:PhysRevLett.51.2124, insightIssue:Spintronics.Nat.Material1, Zutic:2004p477, Awschalom:2007}. Metal/metal interfaces have been extensively studied, and to a lesser degree metal/semiconductor and metal/insulator systems.  However, the transfer of angular momentum between two insulating materials has been more challenging to study due to the lack of suitable detection methods.

NV centers in wide band-gap insulating diamond provide an exceptional platform for performing spin-based measurements.  The paramagnetic NV center is optically active, and its photoluminescence (PL) is dependent on the relative occupation of the lowest lying electronic spin state of the defect center\,\cite{Oort:Glasbeek.JPhysC.1988,Gruber:Wrachtrup.Science.1997}. This enables optical measurement of NV-center spin state with excellent sensitivity, making optically detected magnetic resonance of NV centers an area of intense research activity\,\cite{Balasub:Wrachtrup.Nature.2008, Dutt:Lukin.Science.2007, Fuchs.Awschalom.NatPhys.2011, Grinolds:Yacoby.NatPhys.2013, Kaufman:Wrachtrup.PNAS.2013, Sage:Walworth.Nature.2013, Maletinsky:Yacoby.NatNano.2012, Mamin:Rugar.Science.2013, Maze:Lukin.Nature.2008, Steinert.Wrachtrup.NatComm.2013, Taylor:Lukin.NatPhys.2008, BarGill:Walsworth.NatComm.2013, Toyli:Awschalom.PNAS.2013}.  However, work done thus far relies on manipulation of NV spins using magnetic resonance.



\begin{figure}
\center{\includegraphics[width=1\linewidth]{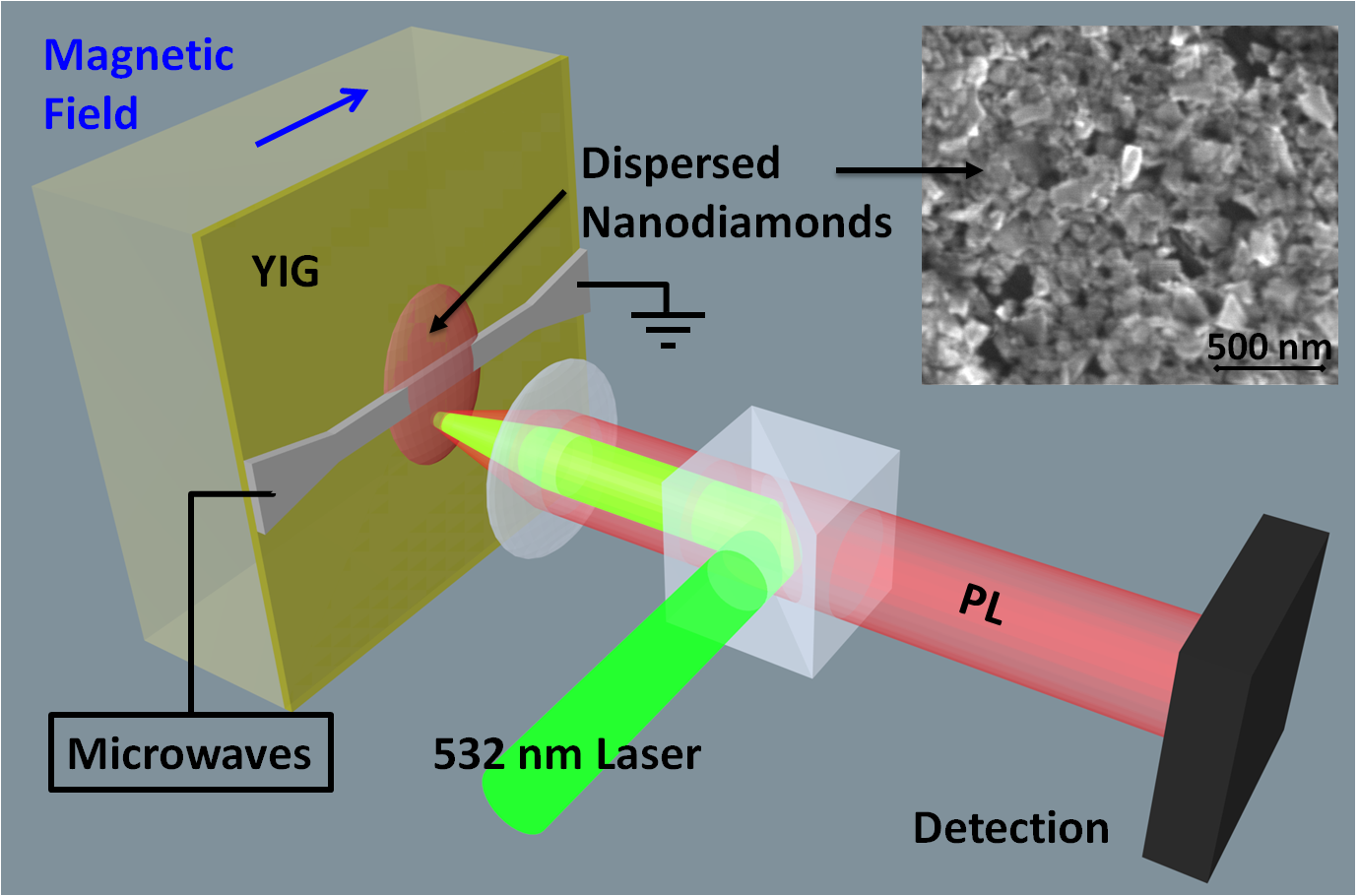}}
\caption{{\bf Experimental Schematic}:  The sample is a 20 nm thick single crystal YIG film with nanodiamonds dispersed on top with a thickness of about 500 nm.  To apply microwave fields to the sample a silver microwire is patterned on the YIG.  Green laser light is focused onto nanodiamonds near the wire, and the intensity of the resulting photoluminescence from the NV centers is measured.  Inset is an SEM image of dispersed nanodiamonds.}
	\label{fig:schematic}
\end{figure}

Here we present experimental evidence that the NV center state can be modified non-resonantly (\emph{i.e.,} by irradiation with microwave magnetic fields at frequencies far from NV center Larmor frequencies) by coupling to the dynamics of a proximal ferromagnetic insulator. This change can be detected as a change in the NV center PL, as is done for conventional microwave-driven resonant spin manipulations. This new effect promises valuable insights into interactions between spins in adjacent dissimilar materials.  This is a particular case in which both materials are insulating and the interaction is effective over long distances ($>300$\,nm).  More generally, this provides a novel method for manipulating NV center spins and could enable sensitive spatially resolved imaging of ferromagnetic phenomena by means atomic scale NV centers\,\cite{Balasub:Wrachtrup.Nature.2008, Grinolds:Yacoby.NatPhys.2013, Maletinsky:Yacoby.NatNano.2012}.

Yttrium Iron Garnet (YIG), Y$_3$Fe$_5$O$_{12}$, was chosen for this experiment as a well known ferrimagnetic insulator with exceptionally low damping\,\cite{Cherepanov.sagaofYIG,Serga:Hillebrands.YIGmagnonics}. Here an epitaxial YIG film, 20nm thick, was grown on a gadolinium gallium garnet (GGG) (111) substrate by off-axis sputtering\,\cite{wang_large_2013, Du_Mechanism_2013, wang_arxiv_2013}. Continuous wave microwave fields are applied to the sample by means of a 300 nm thick and 30 $\mu$m wide silver microstrip line patterned on top of the YIG. Nanodiamonds, 50-200 nm in size (as shown by SEM image analysis) and containing up to a few thousand NV centers each, were dispersed on top of a lithographically defined microstrip line as shown in Fig.~\ref{fig:schematic}. AFM measurement indicates that the nanodiamonds form a 500 nm thick film.    Photoluminescence is excited in the NV centers using a 532 nm laser beam and is collected by a photodiode.  A lock-in measurement is performed on the photodiode signal by modulating the amplitude of the applied microwave field.

\begin{figure}[!]
\center{\includegraphics[width=1\linewidth]{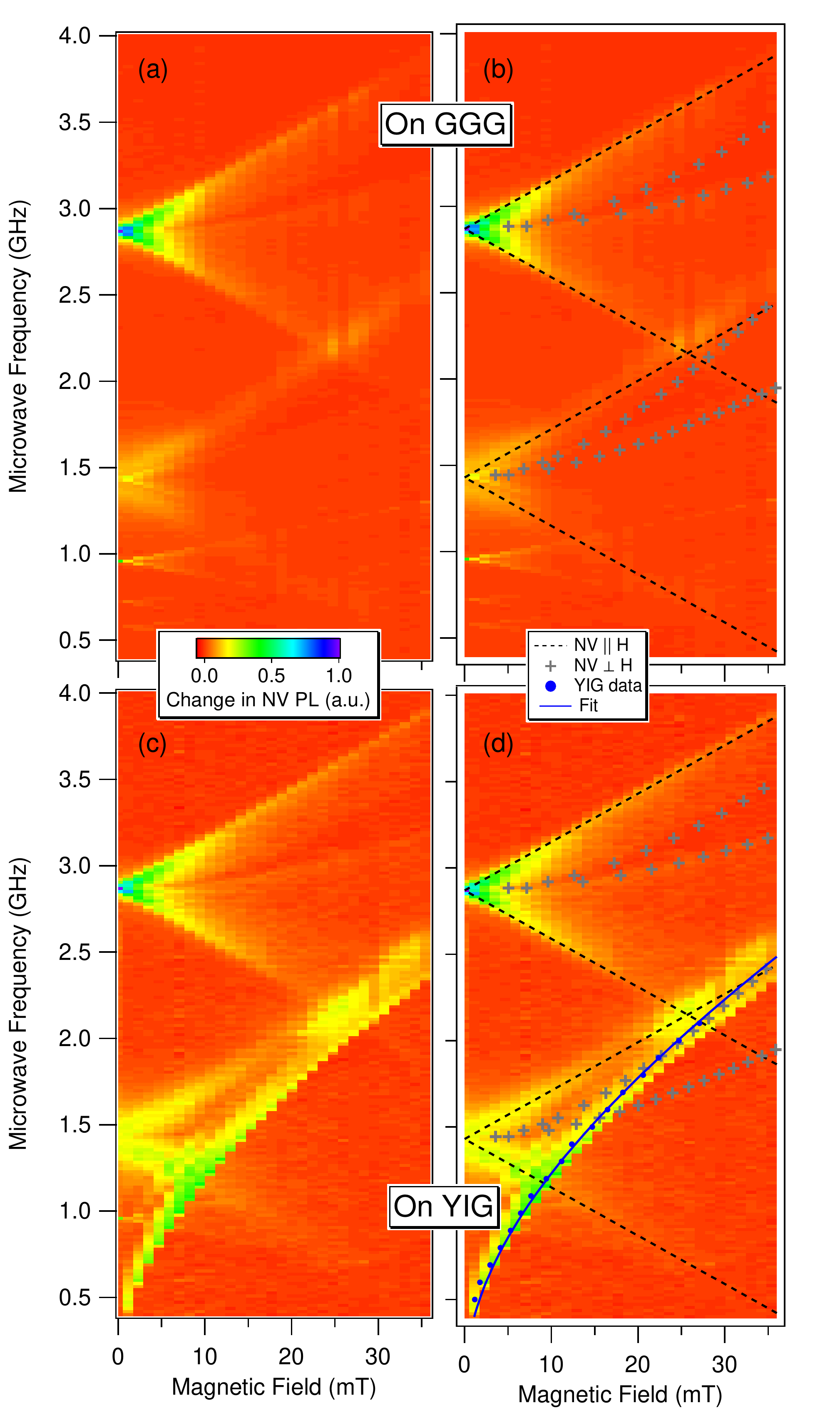}}
\caption{{\bf NV PL Data while YIG undergoes FMR}:  (a) Raw data showing the change in the intensity of the PL from the NV centers in nanodiamonds dispersed on top of GGG, with no YIG, as a function of microwave frequency and magnetic field. (b) The same data overlaid with theoretical resonance conditions for the NV centers.  The black dashed lines show the resonance condition for an NV center with the magnetic field parallel to the NV axis for the ground and excited states.  The grey crosses show the resonance condition for an NV center with the magnetic field perpendicular to the NV axis.  (c) Similar data from nanodiamonds dispersed on the YIG sample with the distinct feature corresponding to the YIG FMR condition.  (d)  The data from (c) with the FMR peaks measured using reflected microwave power (blue dots). Also shown is a fit (blue line) to the calculated dispersion relation for YIG film FMR with the magnetic field in plane (see main text).}
	\label{fig:MainData}
\end{figure}

The lock-in measurement of the resulting modulation of the PL intensity is presented in Fig.~\ref{fig:MainData} as a function of an applied in-plane magnetic field and the applied microwave frequency.  Data for a control sample with nanodiamonds on a GGG substrate without YIG is shown in Fig.~\ref{fig:MainData} (a).  We observe the  intrinsic and well-known magnetic resonances of the NV center ground and excited states, starting at 2.87 GHz and 1.43 GHz respectively.  Shown in Fig.~\ref{fig:MainData} (b) is the same data overlaid with the theoretically expected resonance conditions for the NV centers, which are obtained by solving the NV hamiltonian in the presence of magnetic field parallel and perpendicular to the NV axis (see supplementary information of\,\cite{Balasub:Wrachtrup.Nature.2008}).  We also see several features in the PL below 1.25 GHz (Fig.~\ref{fig:MainData} (a)) that are not related to the normal ground- and excited-state resonances of the NV centers.  These will be discussed in a forthcoming publication.

The data obtained from the nanodiamonds on top of the YIG film is shown in Fig.~\ref{fig:MainData} (c).  The key difference between Fig.~\ref{fig:MainData} (a) and (c) is the feature in (c) that extends up from the lower left-hand corner.  This is highlighted in Fig.~\ref{fig:MainData} (d) where the data from (c) is overlaid with a solid blue curve showing the YIG ferromagnetic resonance condition.  The blue dots show the YIG resonance condition as measured by reflected microwave power. These data are fit (blue curve) using the equation for the uniform FMR mode in a thin film\,\cite{Kittel}, and we obtain a magnetization, $\mu_0M_s$, of 183 mT (see supplementary information for more information).  As can be seen, the intensity of the PL from the NV centers strongly changes precisely when the YIG FMR is excited.

We have considered the effect of heating, caused by the FMR absorption in YIG, on the PL of NV centers. Several control experiments and estimates of the possible effect render this potential explanation highly unlikely.  More details can be found in the supplementary information.

There are two key points to note about the FMR-induced feature in the PL signal.  First, it is seen at frequencies and fields well separated from the NV center's own resonance conditions. This is in clear contrast to the quantum computing and magnetometry techniques being developed, where the spin-state of NV centers is coherently manipulated by microwave fields meeting the magnetic resonance conditions\,\cite{Balasub:Wrachtrup.Nature.2008, Dutt:Lukin.Science.2007, Fuchs.Awschalom.NatPhys.2011, Grinolds:Yacoby.NatPhys.2013, Kaufman:Wrachtrup.PNAS.2013, Sage:Walworth.Nature.2013, Maletinsky:Yacoby.NatNano.2012, Mamin:Rugar.Science.2013, Maze:Lukin.Nature.2008, Steinert.Wrachtrup.NatComm.2013, Taylor:Lukin.NatPhys.2008, BarGill:Walsworth.NatComm.2013}.  Instead, we see a change in the PL that correlates to the excitation of the YIG magnetization into precession by means of ferromagnetic resonance.  It is remarkable that excitations at energies as much as three to six times smaller than any NV center resonance have such a large effect on the NV center spin state. Second, the FMR-induced feature is comparable in amplitude to the intrinsic NV resonances.  The large amplitude of the signal implies that a significant number of NV centers in our laser spot must be contributing to the signal.  This suggests that since the nanodiamond film is ~500 nm thick, the coupling must be either long range (extending hundreds of nanometers) or that spin transport by means of spin diffusion plays a role\,\cite{Ginsparg:1988ui}.

\begin{figure}
\center{\includegraphics[width=1\linewidth]{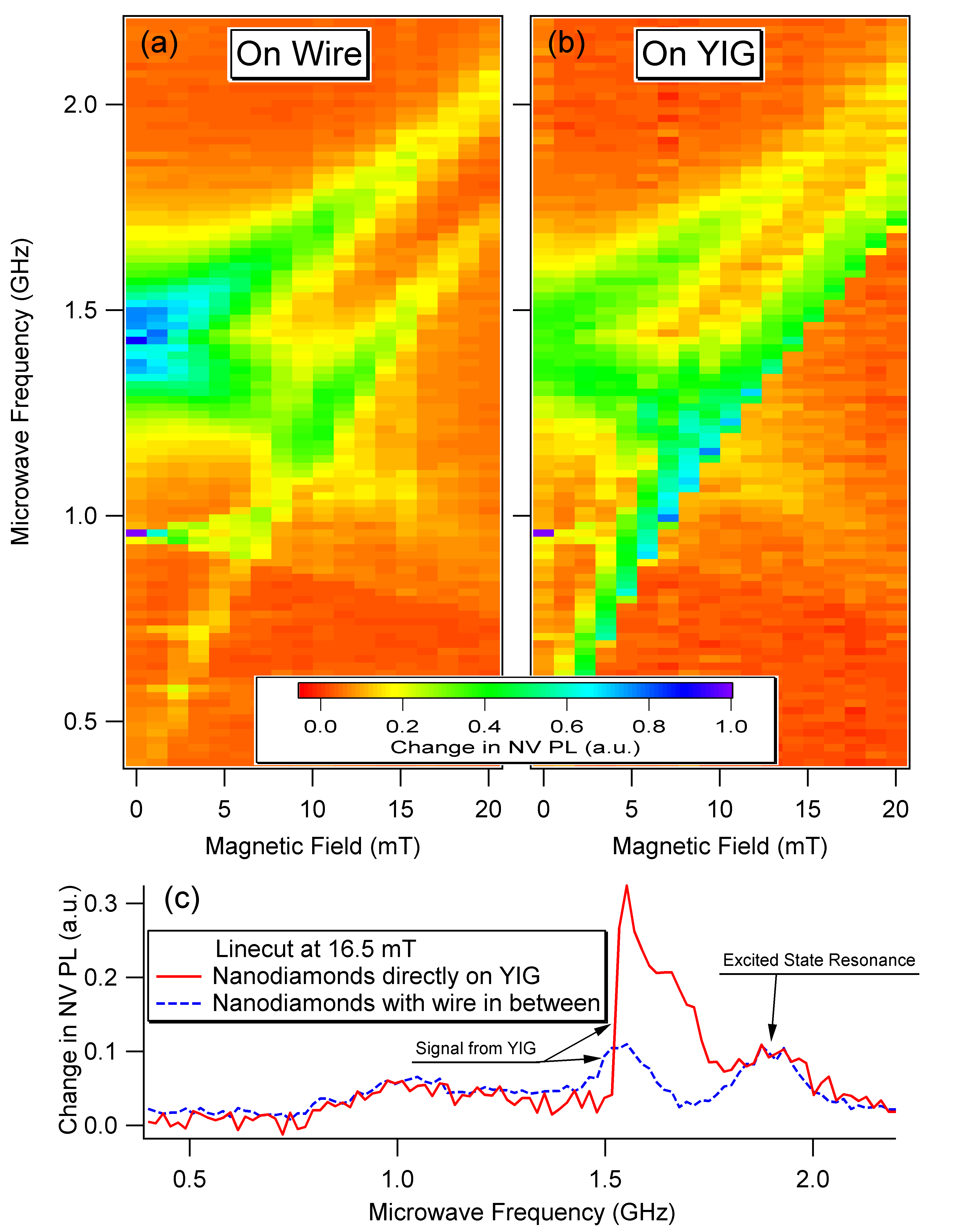}}
\caption{{\bf NV center PL Data on YIG and on Wire Above YIG}:  (a) Data taken with the laser spot focused on top of the $\sim 300$\,nm thick patterned microwire.  The data shows that the signal corresponding to the YIG FMR is still present, though reduced.  The coupling must extend at least 300 nm through the wire.  (b) Data from the nanodiamonds directly on top of the YIG for comparison.  (c) Line cuts of the data in (a) and (b) at 16.5 mT.  The FMR-induced peak is reduced by 3 times relative to the NV excited state peak for the case of nanodiamonds on top of the microstrip, compared to the the case directly on YIG.}
	\label{fig:WireData}
\end{figure}

To probe the spatial extent of the coupling, we repeated the measurement with our laser spot focused on the nanodiamonds on top of the microwire, where the nanodiamonds are separated from the YIG by more than 300 nm.  This data can be seen in Fig.~\ref{fig:WireData} (a) and compared to the signal when the diamond is directly on top of the YIG in Fig.~\ref{fig:WireData} (b).  Linecuts (at 16.5 mT), presented in Fig.~\ref{fig:WireData} (c), show that the FMR-induced feature is reduced but clearly persists even when the nanodiamonds are not in direct contact with the YIG.

While a clear explanation for the effect is not forthcoming from our experimental results, we can make a few observations. First, the insulating nature of both materials rules out long range carrier-mediated transport of spins.  Second, one of the unique aspects of this experiment is that both the YIG magnetization and the NV center spins are out of equilibrium when the YIG is on FMR: YIG due to its resonance and the NV centers due to the hyperpolarizing action of the laser excitation\,\cite{Loubser.RptsProgressPhys.1978, Harrison2004245}. This is in contrast to other systems where angular momentum is transferred between spin sub-systems, e.g. dynamic nuclear polarization or spin pumping\,\cite{slichter, meier_optical_1984, tserkovnyak_enhanced_2002}, where spins in one spin sub-system relax by transferring polarization to another spin sub-system. The transfer of angular momentum from the YIG to the NV centers could result in relaxation towards equilibrium of both the spin systems, and thus be highly desirable for the overall system.

This phenomenon offers the opportunity of probing spin transport in the absence of conductors and employing an all-optical readout. These advantages dramatically reduce the potential confounding effects encountered in studies of spin transport in metallic systems.  Thus this system offers an attractive and powerful approach to better understanding microscopic details of relaxation and spin transport in magnetic heterostructures. Experiments are underway on other ferromagnets and structures to gain more insight on this effect and how angular momentum is transferred between spin systems.

In summary, we have shown that the NV center spin state can be manipulated by a dynamic coupling to the magnetization of YIG.  The availability of a wide selection of ferromagnetic materials and structures (e.g., saturation magnetization and anisotropies) potentially offers a high degree of control in manipulating the NV spin state.  The potential for ultra high resolution imaging of ferromagnetic phenomena using individual NV centers can have a significant impact as well. It should influence the fields of spintronics and quantum information by combining the sensitivity of the NV center and the tunability and scope of ferromagnetism.

The authors wish to thank Yaroslav Tserkovnyak for useful discussions. Funding for this research was provided by the Center for Emergent Materials at the Ohio State University, an NSF MRSEC (Award Number DMR-0820414, ARO (Award number W911NF-12-1-0587) and DOE (Award number DE-FG02-03ER46054).

%

\end{document}